\documentclass[floatfix,pre,twocolumn,superscriptaddress]{revtex4}
\usepackage{latexsym}
\usepackage{graphicx}
\usepackage{amssymb}
\usepackage{amsmath,enumerate,rotate}

\def\CN {\mathcal{N}}

\def\b {\beta}
\def\a {\alpha}

\def\th {\theta}

\def\a {\alpha}

\def\lm {\lambda}

\def\J {\mathbb{I}}

\begin{document} 
\author{F.~Bassetti} \affiliation{Universit\`a degli Studi di Pavia, Dip.
  Matematica, Pavia, Italy } 
\email[e-mail address: ]
{bassetti@dimat.unipv.it}
\author{M.~Cosentino Lagomarsino}
\affiliation{UMR 168 / Institut Curie, 26 rue d'Ulm 75005 Paris, France}
\affiliation{Universit\`a degli Studi di Milano, Dip.
    Fisica, Via Celoria 16, 20133 Milano, Italy } 
\email[ e-mail address: ]{mcl@curie.fr}
\author{B.~Bassetti} 
\affiliation{Universit\`a degli Studi di Milano, Dip.
    Fisica, Via Celoria 16, 20133 Milano, Italy } 
\affiliation{I.N.F.N., Milano, Italy} 
\email[e-mail address: ]{bassetti@mi.infn.it}
\author{P.~Jona} 
\affiliation{Politecnico di Milano, Dip. Fisica, Pza Leonardo Da Vinci
  32, 20133 Milano, Italy} 

\date{\today}

\title{Random Networks Tossing Biased Coins}

\begin{abstract}
  In statistical mechanical investigations on complex networks, it is useful
  to employ random graphs ensembles as null models, to compare with
  experimental realizations. Motivated by transcription networks, we present
  here a simple way to generate an ensemble of random directed graphs with,
  asymptotically, scale-free outdegree and compact indegree. Entries in each
  row of the adjacency matrix are set to be zero or one according to the toss
  of a biased coin, with a chosen probability distribution for the biases.
  This defines a quick and simple algorithm, which yields good results already
  for graphs of size $n \sim 100$. Perhaps more importantly, many of the
  relevant observables are accessible analytically, improving upon previous
  estimates for similar graphs.  The technique is easily generalizable to
  different kinds of graphs.
\end{abstract}

\maketitle

\section{Introduction.}
In our investigation concerning transcription networks, we came across a
simple and effective way to generate a random ensemble of directed graphs
having similar features as the experimental ones.  Transcription networks are
directed graphs that represent regulatory interactions between genes.
Specifically, the link $a \rightarrow b$ exists if the protein coded by gene
$a$ affects the transcription of gene $b$ in mRNA form by binding along DNA in
a site upstream of its coding region~\cite{BLA+04}.
For a few organisms, such as E.~coli and S.~cerevisiae, a significant fraction
of the wiring diagram of this network is
known~\cite{LRR+02,GBB+02,SSG+01,HGL+04}.
%
The hope is to study these graphs, together with the available information on
the genes and the physics/chemistry of their interactions, to infer
information on the large-scale architecture and evolution of gene regulation
in living systems.
In this program, the simplest approach to take is to study the
topology of the interaction networks.
For instance, order parameters such as the connectivity and the clustering
coefficient have been considered~\cite{GBB+02}.

To assess a topological feature of a network, one typically generates so
called ``randomized counterparts'' of the original data set as a null model.
That is, an ensemble of random networks which bare some resemblance to the
original.
The idea behind it is to establish when and to what extent the observed
biological topology, and thus loosely the living system under exam, deviates
from the ``typical case'' statistics of the null ensemble.  For example, a
topological feature that has lead to relevant biological findings, in
particular for transcription, is the occurrence of small subgraphs - or
``network motifs''~\cite{MSI+02,MIK+04,MBV05,WA03,YSK+04}.
The choice of what feature to conserve (or not) in the randomized counterpart
is quite delicate and depends on specific considerations on the
system~\cite{IMK+03}.  The null ensemble used to discover motifs usually
conserves the degree sequences of the original network, that is, the number of
regulators and targets of each node.
The observed degree sequences for the known transcription networks follow a
scale-free distribution for the outdegree, with exponent between one and two,
while being Poissonian in the indegree~\cite{GBB+02,CLB05}.
Motifs are then interpreted as elementary circuit-like building blocks and
have been shown in many cases to work independently~\cite{MA03}.
  \begin{figure}[tb]
    \centering \includegraphics[width=.33\textwidth]{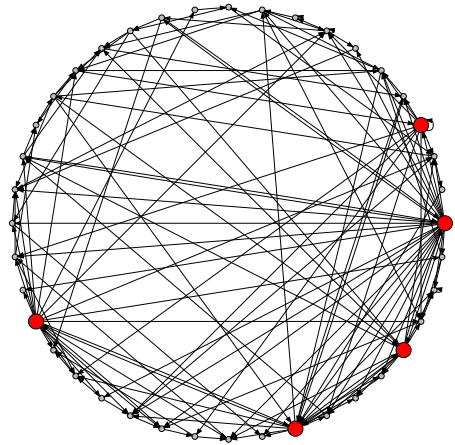}
    \caption{Example of a graph generated with our algorithm with $n=40,
    \ \beta=2.8, \ \alpha=1$. Nodes with more than ten outgoing edges are
    larger (red online).}
    \label{fig:grafofig}
  \end{figure}
In connection with this line of research, it is interesting to study random
ensembles of graphs with probability distributions for the degree sequences
that are similar to those observed experimentally, with the objective of
characterizing theoretically some relevant topological observables, such as
the subgraph distributions~\cite{IMK+03,BB03}.
Here, we describe a simple, and fast, algorithm that performs this
task by tossing coins with prescribed random biases.  Differently from
more sophisticated techniques available in the
literature~\cite{IMK+03,RJB96,CDH+05,MKI+03,MR95}, our method is not
designed to conserve a degree sequence, but rather as a general random
graph model, that, in particular, can be used to generate graphs with
degree distributions that agree with the observed power-law
distributed outdegrees and compact indegrees~\cite{IMK+03}. 
To this aim, the ensemble will be generated by a parametric model,
where the adjustable parameters can be used for fits of real
data-sets. Note that, with the weaker constraint on the degree
distribution that we have chosen, it would be very inconvenient to
generate the ensemble throwing degree sequences \emph{a priori} from
the given distributions and then using an algorithm designed for fixed
degree sequences, which is necessarily more costly.
We will see that, because of the extreme simplicity of our
formulation, some observables can be computed analytically rather than
estimated as in ref.~\cite{IMK+03}.
After introducing the algorithm and showing that the ensemble has the required
features, we will compute the number of some observables that are relevant for
transcription, such as triangular subgraphs.

\section{Algorithm.} 

\begin{figure}[htb]
  \includegraphics[width=.36\textwidth]{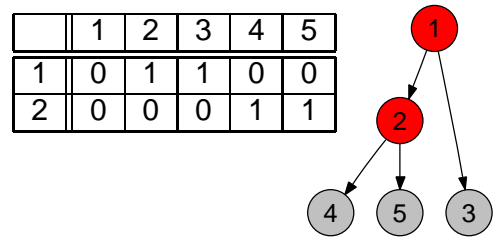}
  \caption{Example of a rectangular matrix and its associated graph.
    Nodes 1 and 2 represent transcription factors, and can regulate
    any other node. Nodes 3 to 5 are targets and only receive incoming
    links. In this case $m_n = 2/5 n$. }
  \label{fig:delta}
\end{figure}

Any directed graph $G_n$ with $n$ nodes is completely described by its
adjacency matrix $ A(G_n)=[x^{(n)}_{i,j}]_{i,j=1,\dots, n} $, where
$x^{(n)}_{i,j} = 1$ if there is a directed edge $i \rightarrow j$, $0$
otherwise.  Instead of square matrices, one may also consider rectangular
matrices with a prescription on the scaling of the rows with the columns.  In
what follows we will deal with rectangular matrices $m_n \times n$ with $m_n
\leq n$. As we will see, this is particularly useful for networks with
power-law degree distributions having exponent equal or lower than two (for
which the average diverges), to keep the asymptotics well-behaved.  In the
context of transcription networks, the hypothesis of rectangularity is
well-motivated by the fact that typically only a subset of $m_n$ nodes encode
for transcription factors (namely, they have outgoing edges). Thus, in a $m_n
\times n$ matrix, the first $m_n$ columns will contain the incoming links to
the transcription factors, and the next $n-m_n$ columns will correspond to non
transcription-factor encoding genes. Note that in general nodes that send out
edges are also receiving edges (Fig~\ref{fig:delta}).

Our model ensemble can be defined by the following generative algorithm. For
each row of $A$, (i) throw a bias $\theta$ from a prescribed probability
distribution $\pi_n$ (ii) set the row elements of $A$ to be $0$ or $1$
according to the toss of a coin with bias $\theta$.  Since each row is thrown
independently, the resulting probability law is
\begin{equation}
  \begin{array}{l}
    P\{ x_{i,j}^{(n)} = e_{i,j}, \, i=1,\dots,m_n, j=1,\dots,n \}= \\
     \, \\
    \prod_{i=1}^{m_n} \int_{[0,1]} \th_i^{\sum_{j=1}^n
      e_{i,j}}(1-\th_i)^{n-\sum_{j=1}^n e_{i,j}} \pi_n(d\th_i)   
  \end{array}
  \label{eq:P}
\end{equation}
where $e_{i,j} \in \{0,1\}$, $i=1,\dots,m_n$, $j=1,\dots n$. Note that the row
elements are not independent~\cite{note1}. Eq.~(\ref{eq:P}) is a general
probability distribution based on two symmetries: (a) the fact that a node
regulates other ones is independent from the nodes regulated by other genes
(b) the identity of the regulated nodes is unimportant, and what matters is
their number only.
The two symmetries can be summarized by saying that the indegree and
the outdegree are uncorrelated~\cite{MS05,New02}.
It is worth noticing that our model could also be seen as a special
case of a directed graph variant of the so called hidden variables
models, introduced in ~\cite{CCLRM2002}, see also
~\cite{BPS2003,S2006}.  In this very general class of undirected
random graphs the quantity $\th$ is interpreted as the "fitness" of
each vertex and the emphasis is on the problem of how power-laws may
emerge "naturally" in interaction networks.
To complete our model, one has to specify the choice for $\pi_n$,
which determines the behavior of the graph ensemble. We choose the
two-parameter distribution 
\begin{equation}\label{defpi}
\pi_n(d\th)=  Z^{-1} \th^{-\beta}\chi_{(\frac{\a}{n},1]}(\th) d\th
\end{equation}
where $\a >0$ and $\beta >1$ are free parameters, $\chi_{(\frac{\a}{n},1]}$ is
the characteristic function of the interval $(\frac{\a}{n},1]$, taking the
value one inside the interval and zero everywhere else, and $
Z:=\frac{(n/\a)^{\beta-1}-1}{\beta-1} $ is the normalization constant.
In simple words, Eq.~\eqref{defpi} defines the probability to take a coin with
a certain bias $\theta$, which is connected to the outdegree of the
corresponding node.
As we will see, the function $\th^{-\beta}$ gives
a power-law tail to the outdegree.  Conversely, the cutoff on $\th$ defined by
$\alpha$ poses a constraint on the number of low outdegree nodes, and will be
used to control the indegree distribution.
In concrete applications at finite sizes, it might be useful to
introduce also an upper cutoff on $\pi_n$, that is
\begin{equation}\label{defpi2}
  \pi_n(d\th) \propto Z^{-1}
  \th^{-\beta}\chi_{(\frac{\a}{n},1-\frac{\gamma}{n}]}(\th) d\th.  
\end{equation}
This does not affect the asymptotic results given below but gives more
flexibility to the model.  Hence, in what follows, with the exception
of Section \ref{sec:fit}, we shall take $\gamma=0$.

\section{Results.}

An example of a graph generated with our algorithm is shown in
Fig~\ref{fig:grafofig}.
The algorithm is quite efficient: its computational cost is determined
by the number of coin tosses (each of which takes the same amount of
operations) and thus scales like $n^{2}$.  Our Fortran 77
implementation running under Linux on an AMD Athlon 64 X2 3800+ PC,
generates a graph with $n=10^{4}$ in about $3.5$ seconds.
Many observables can be computed knowing the probability of the link $i
\rightarrow j$, $\mu_n:=P\{x_{i,j}^{(n)}=1\} = \int_{[0,1]} \th \pi_n(d \th)$.
By simple calculation from Eq.~(\ref{eq:P}) and~\eqref{defpi}, we get
\begin{displaymath}
\mu_n = 
\left\{%
\begin{array}{ll}
\frac{(\beta-1)\a^{\beta-1}}{(2-\beta)n^{\beta-1}}\frac{
1-\left(\frac{\a}{n} \right)^{2-\beta}}{1-\left (\frac{\a}{n}
\right)^{\beta-1}}  \quad \mathrm{if}  \ \    1 < \beta < 2 \\
\frac{\a}{n-\a}(\log n-\log \a)  \quad \mathrm{if} \ \ \beta=2 \\
\frac{(\beta-1)}{(\beta-2)}\frac{ \left (\frac{n}{\a}
\right)^{\beta-2}-1 }{\left (\frac{n}{\a} \right)^{\beta-1}-1 } \quad
\mathrm{if} \ \ \beta >2 \\
\end{array}%
\right.
\end{displaymath}
Note that the formulas above for $\b>2$ and $\b<2$ are identical, but have
been recast to show the leading terms in the scaling.  Hence $\mu_n$, for $n
\to \infty$, scales as $1/n^{\beta -1}$ if $1 < \beta < 2$, as $(\log n)/ n$
if $\beta=2$, and as $1/n$ if $\beta >2$.  Note that $\mu_n$ is directly
related to the mean number of links in the network, which can thus be
controlled through the parametric dependency of this quantity.
We did not prove anything regarding the emergence of a giant component. The
graphs we generated numerically seem to have a large component. On the other
hand, analytically, it is not hard to see that probability that a graph $G_n$
generated with our technique has only one connected component goes to zero as
$n$ diverges.


\subsection{Degree Distributions.}

The variable $Z_{m_n,j}:=\sum_{i=1}^{m_n} x^{(n)}_{i,j} $ represents the
indegree of the $j$-th node in the random graph, while $S_{n,i}:=\sum_{j=1}^n
x^{(n)}_{i,j} $ represents the outdegree of the $i$-th node ($1 \leq i \leq
m_n$). Clearly, the mean degrees are equal to $m_n \mu_n$ and $n \mu_n $,
respectively.  To access the degree distributions, one has to compute $
P\{S_{n,i}=k\}= {{n}\choose{k}}\int_{[0,1]} \th^k(1-\th)^{n-k} \pi_n(d \th) $
and $ P\{Z_{m_n,j}=k\}= {{m_n}\choose{k}} \mu_{n}^k(1-\mu_n)^{m_n-k} $.
\begin{figure}[htb]
  \includegraphics[width=.36\textwidth]{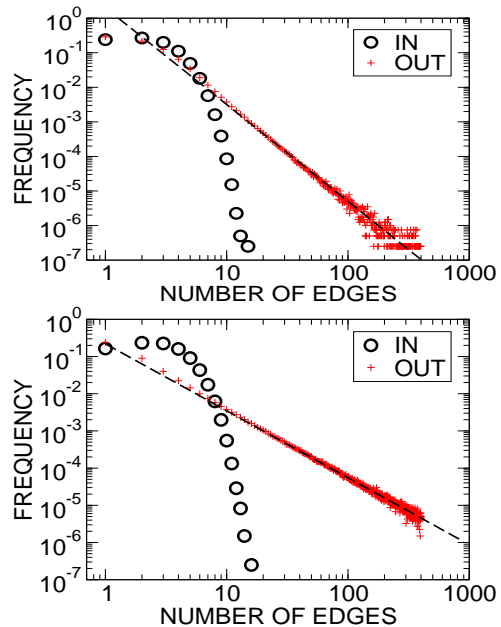}
  \caption{Degree distributions (in logarithmic scale) of two graphs generated
    with our algorithm. The two panels correspond to graphs having $n = 400$,
    square adjacency matrices and different values of the parameters.  Top:
    $\beta=2.8, \ \alpha=1$.  Bottom: $\beta=1.8, \ \alpha=0.2$. To obtain a
    compact indegree distribution in the case of $\beta \le 2$ one has to
    supply smaller values of $\alpha$. The dashed lines are power-laws with
    exponent $\beta$.}
  \label{fig:degree}
\end{figure}

Let us concentrate first on the outdegree. An evaluation of its
distribution yields the following asymptotic law for $n \to \infty$ for
any $\a >0$ and $\b >1$:
\begin{displaymath}
  P\{S_{n,j}=k \} \sim p_{\a,\b}(k) =   \frac{\a^{\b-1}(\b-1)}{k!} 
  \int_{\a}^{+\infty} t^{k-\b}e^{-t} dt.
\end{displaymath}
It is easy to show that $p_{\a,\b}(k)$ has a power-law tail.  Indeed, if $k >
\b$, $p_{\a,\b}(k)=\a^{\b-1}(\b-1)
(\frac{\Gamma(k+1-\b)}{\Gamma(k+1)}-\frac{1}{\Gamma(k+1)}\int_0^{\a}
t^{k-\b}e^{-t} dt)$ (where $\Gamma$ indicates the gamma function). 
Thus, since $\frac{\Gamma(k+1-\b)}{\Gamma(k+1)} \sim \frac{1}{k^\b}$, one
concludes that
$$
p_{\a,\b}(k) = \frac{1}{k^\b}(\a^{\b-1}(\b-1)+o(1)) \ \ .
$$
Fig.\ref{fig:degree} shows the degree distributions of numerically
generated examples for $n=400$.  In practice, already at $n \sim 100$ one gets
a very marked power-law in the tail of the outdegree distribution.
Considering now the indegree, since its behavior is determined by $\mu_n$, one
has to distinguish among the different possible scalings for this quantity.
The simplest case is $\b >2$, where for $m_n=]\delta n[$ ($\delta$ being any
constant in $(0,1]$ and $]x[$ being the integer part of $x$) and for $n \to
\infty $, using the Poissonian approximation of a binomial distribution, it is
immediate to show that $ P\{Z_{m_n,j}=k \} \sim \frac{e^{-\lm} \lm^k}{k!}  $,
with $\lm=\frac{\delta \a (\beta-1)}{(\beta-2)}.$ Things are slightly more
complicated for $\b \le 2$. Here, essentially because of the scaling for
$\mu_n$ in the limit $n \to \infty$, the indegree distribution diverges if one
chooses $m_n=]\delta n[$. Thus, to obtain a well-behaved asymptotic
distribution, one has to compensate more strongly for the scaling of $\mu_n$
with the number of rows of $A$.  For $\b = 2$, the necessary choice is $
m_n=]\delta n/\log n[$ rows, and for $1<\b<2$ one has to take $m_n=]\delta
n^{\b-1}[$ rows.
With these prescriptions, the indegree distribution is asymptotically Poisson,
and has the form $\frac{e^{-\lambda} \lambda^k}{k!}$ with $\lambda=\delta \a$,
or $\lambda=\delta \a^{\b-1}\frac{\b-1}{2-\b}$, for $\b = 2$ and $1<\b<2$
respectively.  In other words, asking for a degree distribution that brings to
an outdegree having a power-law tail with divergent mean ($\beta \le 2$) poses
a heavy constraint on the number of regulator nodes (the rows of the matrix).
On the other hand, for the purpose of generating square ($n \times n$)
matrices at finite $n$ with $\b \le 2$ and compact indegree, this issue is not
so important. A suitable choice of the parameter $\alpha$ (see
Fig.~\ref{fig:degree}) can solve the problem.  In what follows we will discuss
mainly the case of square matrices.

\subsection{Subgraphs.}

The simple structure of the random graphs generated by our algorithm makes it
possible to compute analytically the mean value of the number of subgraphs of
a given shape contained in the graph.  Consider a subgraph $H$, with $k$ nodes
and $m$ edges, that is, the set of ordered pairs of nodes $ H=\{ i_1 \to
i_{1,1}, \dots, i_1 \to i_{1,m_1}, i_2 \to i_{2,1}, \dots, i_k \to i_{k,1},
\dots, i_k \to i_{k,m_k} \} $, where $\sum_{i=1}^k m_i=m$.  For example, $ i_1
\to i_2, i_2 \to i_3, i_3 \to i_1 $ denotes a ``feedback loop''
(\texttt{fbl}), or a 3-cycle.  Now, if $G_n$ is a random graph with $n$ nodes
generated by our algorithm, the probability to observe $H$ as a subgraph of
$G_n$ can be written as
\[
P\{H \in G_n\}= \int_{[0,1]} \th_1^{m_1} \pi(d\th_1)
 \dots  \int_{[0,1]}
\th_k^{m_k} \pi_n(d \th_k) .
\]
To compute the mean of the number $\CN_H(G_n)$ of subgraphs isomorphic to $H$
one also has to consider the quantity $N(H)$ of subgraphs isomorphic to $H$
contained in the complete graph with $n$ nodes. The desired average is then
$<\CN_H(G_n)> =N(H) P\{H \in G_n\}$ (where $<..>$ denotes the mean).  Things
are slightly more complicated for rectangular matrices because in the
evaluation of $N(H)$ one needs to take into consideration also the constrains
given by the fact that only $m_n$ nodes can have out edges.

\begin{figure}[htbp]
  \centering
  \includegraphics[width=.45\textwidth]{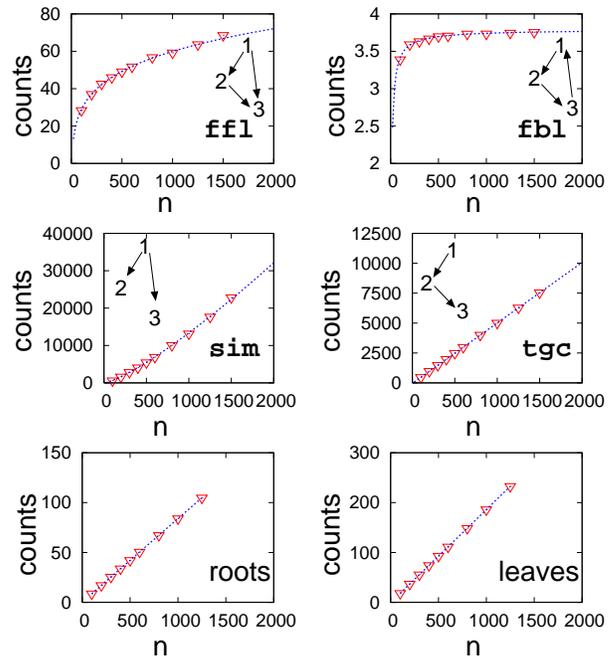}
  \caption{Comparison between analytical (dotted lines) and numerical
    (triangles) evaluations of the mean number of some observables as a
    function of system size $n$, for $\beta=2.8, \ \a=1$. Numerical averages
    are evaluated on $10^{5}$ realizations. Top and middle: mean number of
    three-node subgraphs. Each subgraph is sketched next to its corresponding
    plot.  Top: feedforward and feedback loops (\texttt{ffl} and
    \texttt{fbl}).  Middle: two kinds of open triangles, that can be termed
    ``single input modules'' (\texttt{sim}) and ``three-gene chains''
    (\texttt{tgc}). Bottom: roots and leaves.}
  \label{fig:motifs}
\end{figure}

As an example, we evaluate now, in the case of square matrices, the mean
number of feedback loops versus feedforward loops, which play an important
role for transcription~\cite{MA03}.
A feedforward loop (\texttt{ffl}) is a triangle with the form $i_1 \to i_2 \to
i_3, i_1 \to i_3$. It is found to be a motif in known transcription networks,
and identified with the function of persistence filter or amplifier.
Conversely, feedback loops (which in principle could form switches and
oscillators) are usually not found in transcription
networks~\cite{SMM+02,MIK+04}. Following the procedure described above, one
gets $ <\CN_{\mathtt{fbl}}(G_n)>=2 {n \choose 3}\mu_n^3$ (this holds also for
$k$-cycles, with $k$ in place of 3).  Once more, this can be evaluated
analytically with straightforward calculations. As it depends only on the
behavior of $\mu_n$, its scaling for large $n$ easily follows.  The evaluation
of feedforward loops is slightly more complicated. In general
\begin{displaymath}
  <\CN_\mathtt{ffl}(G_n)> =6 {n \choose 3} \int_{[0,1]} \th^2 \pi_n(d\th)
  \int_{[0,1]} \th \pi_n(d\th),
\end{displaymath} 
and hence, under (\ref{defpi}),
\begin{displaymath}
  \begin{array}{lll}
<\CN_\mathtt{ffl}(G_n)> &=& 6 {n \choose 3}
\frac{(\b-1)^2}{[(n/\a)^{\b-1}-1]^2 } \times \\
\, \\
&&\int_{\a/n}^1 \th^{2-\b} d\th \int_{\a/n}^1 \th^{1-\b} d\th.
\end{array}
\end{displaymath}
Note that the finite $n$ formulas above can be computed explicitly, and so
does their scaling for finite sizes.  In appendix~\ref{sec:fflav}, we spelled
out the example of \texttt{ffl}s to exemplify this point.

In Fig.~\ref{fig:motifs}, we report a comparison of the exact calculation of
some triangular subgraphs with results obtained from numerical evaluation. The
agreement between the analytical expressions and the numerics is perfect.
Having analytically exact expressions for any system size can be an advantage
with respect to models where only asymptotically exact expressions are
available, especially thinking that many concrete datasets have relatively
small sizes.  Moreover, it is possible to compute analytically the standard
deviation of the number of subgraphs. For example, we considered again the
number of feedback loops and feedforward loops. The most interesting fact is
that for $\beta > 2$, the former are always more widely distributed.  A sketch
of the calculation and some results are reported in Appendix \ref{sec:A1}.
 
Finally, one can evaluate the scaling behavior of the ratio of feedback and
feedforward loops, which is given below
\[
\frac{\langle\CN_{\mathtt{ffl}}(G_n) \rangle}{\langle\CN_{\mathtt{fbl}}(G_n)\rangle}
 \sim 
\left\{%
\begin{array}{ll}
n^{\b-1}  &\,  \, \,  \,\mathrm{if} \ \  1 < \b < 2  \\
n/(\log n)^2 &\,  \, \,  \,\mathrm{if} \ \ \b = 2 \\
    n^{3-\b} &  \,  \, \,  \mathrm{if} \ \  2 < \b < 3 \\
 \log n &  \,  \,  \,\mathrm{if} \ \  \b =3 \\
  \lm &  \,  \, \,\mathrm{if} \ \ \b > 3 \\
\end{array}%
\right.
\]
where $\lm =3(\beta-2)^2(\b-3)^{-1}(\b-1)^{-1} > 1$.  Thus, the
\texttt{ffl} always dominate, although there is a wide range of
regimes. Note that the dominance of feedforward triangles is even
stronger if one considers the rectangular adjacency matrices discussed
above. For example, for $1 < \b < 2$, and rectangular matrices, we
calculate $ \frac{ \langle\CN_{ \mathtt{ffl}}(G_n) \rangle}{\langle
  \CN_{\mathtt{fbl}}(G_n)\rangle} \sim n$.

\subsection{Roots and Leaves.}
As a second example, we report the calculation of the mean number of roots
(nodes with only outgoing links) versus leaves (nodes with only incoming
links). More specifically, we say that $i$ is a root if there is no edge of
the kind $j \to i$, but there is at least one edge of the kind $i \to j$, with
$j \not = i$.  Loops do not count. Conversely, we say that $i$ is a leaf if
there is no edge of the kind $i \to j$, but there is at least one edge of the
kind $j \to i$, with $j \not = i$.  Again we exclude loops and isolated
points. We find the following scaling for the numbers
of roots $\mathfrak{R}$, and of leaves~$\mathfrak{L}$:
\[
<  \mathfrak{L}(G_n) > \sim n
\]
while
\[
<  \mathfrak{R} (G_n) > \sim
\left\{%
\begin{array}{ll}
    n &  \,  \,\mathrm{if} \ \ \b > 2\\
    n^{1-\a} &  \,  \,\mathrm{if} \ \ \b = 2  \\
  e^{-\tau^2 n^{2-\beta}}  &\,  \, \mathrm{if} \ \ 1 < \b < 2  \\
\end{array}%
\right.
\]
where $\tau^2=\frac{\b-1}{2-\b} \a^{\b-1}$.
Once again, we stress that these quantities are accessible analytically, and
there is perfect agreement between the data generated by the algorithm and the
calculations.

\subsection{Hub.}  
As a last example of important observable in our graph ensemble, we discuss
the distribution and mean number of hubs.  The so--called hub is the node
having maximal outdegree among the nodes, that is,
$H_n:=\max_{i=1,\dots,m_n}(S_{n,i})$. Once again, it is possible to give an
analytical expression of the limit law of the hub under a suitable rescaling.
Indeed, by stochastic independence, it is clear that $P\{ H_n \leq x b_n\}=
(1- P\{S_{n,i} > x b_n \})^{m_n}$, where $x>0$ is any positive number.
Moreover, it is not too hard to prove that, for suitable choices of $b_n$ and
$m_n$, $P\{S_{n,i} > x b_n \}=1/m_n[(\a/x)^{\b-1}+o(1)]$.  More precisely, for
$\b \geq 2$ and for every positive number $x$
\[
P\{ H_n/b_n  \leq x \} \sim e^{-(\a/x)^{\b-1}}.
\] 
The above expression gives the effective probability distribution that one can
use for the hub outdegree in the asymptotic limit.  In particular, for $\b
>2$, $m_n=n$ and $b_n=n^{1/(\b-1)}$, and, with some work, we prove that $<H_n>
\sim n^{1/(\b-1)}$, as found in~\cite{IMK+03}.  For $\b=2$, one has to take
$m_n=b_n=n/\log n$, which lead to analogous scaling results.  
Finally, for $1
< \b < 2$ and $m_n=n^{\b-1}$, one gets the expression
\[
P\{ H_n/n \leq  x \}  \sim
e^{-(\a/x)^{\b-1}}\chi_{(0,1)}(x)+\chi_{[1,\infty)}(x) 
\]
for every positive $x$. Note that in this last case the probability of finding
a hub of size $n$ is asymptotically finite, and equal to $1-
e^{-(\a)^{\b-1}}$. This concentration effect was already noted
in~\cite{IMK+03} using a different random graph model, without computing
explicitly the asymptotic probability distribution.
It is worth recalling that $e^{-(\a/x)^{\b-1}}\J_{[0,+\infty)}(x)$ is
the Frechet type II extreme value distribution, i.e. one of the three
kinds of extreme value distributions that can arise from limit law of
maximum of independent and identically distributed random variables
(see for instance \cite{galambos}). For extreme values distributions in
scale-free network models see, e.g., \cite{moreira}.


\section{Other Applications.}
\label{sec:var}

While here we restricted our attention to the case of directed graphs
with compact indegree and power-law outdegree, our coin-toss method of
generating exchangeable graphs is more general and has a wider range
of application. For example, one can consider the following algorithm:
(i) throw a bias $\theta$ from a prescribed probability distribution
$\pi_n$ (ii) set all the elements of $A$ to be $0$ or $1$ according to
the toss of a coin with bias $\theta$.  The resulting probability law,
for square matrices, is
\[
\begin{split}
Q\{ x_{i,j} = e_{i,j};& \, i,j=1,\dots,n \}= \\ &\int_{[0,1]}
\th^{\sum_{i,j} e_{i,j}}(1-\th)^{n^2-\sum_{i,j} e_{i,j}} \pi_n(d\th) \\
\end{split}
\]
$e_{i,j}$ being any element in $\{0,1\}$ $i,j=1,\dots n$.  Again set
$\mu_n:=Q\{x_{i,j}^{(n)}=1\} = \int_{[0,1]} \th \pi_n(d \th)$.  The
resulting ensemble of random graph has a large variability in the
number of links. In the $n\times n$ case, the degree distributions are
given by 
\( Q\{S_{n,i}=k\}= Q\{Z_{n,j}=k\}
={{n}\choose{k}}\int_{[0,1]} \th^k(1-\th)^{n-k} \pi_n(d\th).  \) 
Assuming (\ref{defpi}) one gets
\[
 Q\{S_{n,j}=k \} \sim  Q\{Z_{n,i}=k\} \sim p_{\a,\b}(k),
\]
which has, again, a power-law tail. For this model, quantities like the mean
number of subgraphs, roots, leaves and hubs, are again easily computed
analytically, in the same way we described above.  
Furthermore, throwing a triangular matrix with the same algorithm, one can
easily generate a power-law model for undirected graphs.  Finally, variants of
the model can be generated by changing the probability distribution $\pi_n$
for the biases.  Overall, all these possibilities remain open to explore and
could be useful to generate both analytically solvable random graph models and
quicker algorithms in many applications.


\section{Example of Comparison with Empirical Data.}
\label{sec:fit}

A detailed comparison between known real transcriptional networks and
the null models obtained with our coin-toss algorithm is beyond our
aims here.
Nevertheless, to show that our model can be used for direct
statistical comparisons, as an alternative to the more stringent
constraint of preserved degree sequences, we present here a brief
application to the Shen-Orr~\cite{SMM+02} dataset for the E.~coli
Transcription Network.
Motifs discovery, for example, entails comparing the occurrence of subgraphs
in a real network with a null ensemble. This null ensemble can be obtained
from our coin-toss model, with some prescribed parameter set. The parameters
can be chosen by performing a fit of the model graphs with some observed
features of the data, such as, for example, decay of the degree distributions
and number of regulatory elements (additional parameters can also be
introduced if needed).  The chosen ``invariants'' can be motivated
biologically.

We generated our random ensemble with distribution $\pi_n$ given by
(\ref{defpi2}) as follows.  First, from a frequentistic estimate of
$\pi_n$ we determined the probable value of $\beta$ and of the cutoff
on the maximum, i.e. $\gamma$.  This last quantity has to be regarded
as a biological constraint, and is necessary to obtain an ensemble
having on average the same number of links as the empirical one; we
measure the upper cutoff $1- \gamma /n$ to be about $18\%$. The
estimated value for $\beta$ ranges from 1.6 to 2.1, depending on the
binning of the histogram of $\pi_n(\theta)$. We note that these values
are larger than those that obtained from fitting direcly on the
outdegree sequence. As a second step, we fixed the rectangularity of
the matrix with the ratio of transcription factors to total number of
nodes, choosing $\delta$ such that $m_n/n \simeq 0.2766$.  Finally, we
fitted $\alpha$ to reproduce on average the observed number of links
and nodes.  In practice, since the model nturally produces a certain
number of isolated nodes, one has to generate slightly larger matrices
and compare the submatrix made of non-isolated nodes.

\begin{figure*}[htb]
  \includegraphics[width=.8\textwidth]{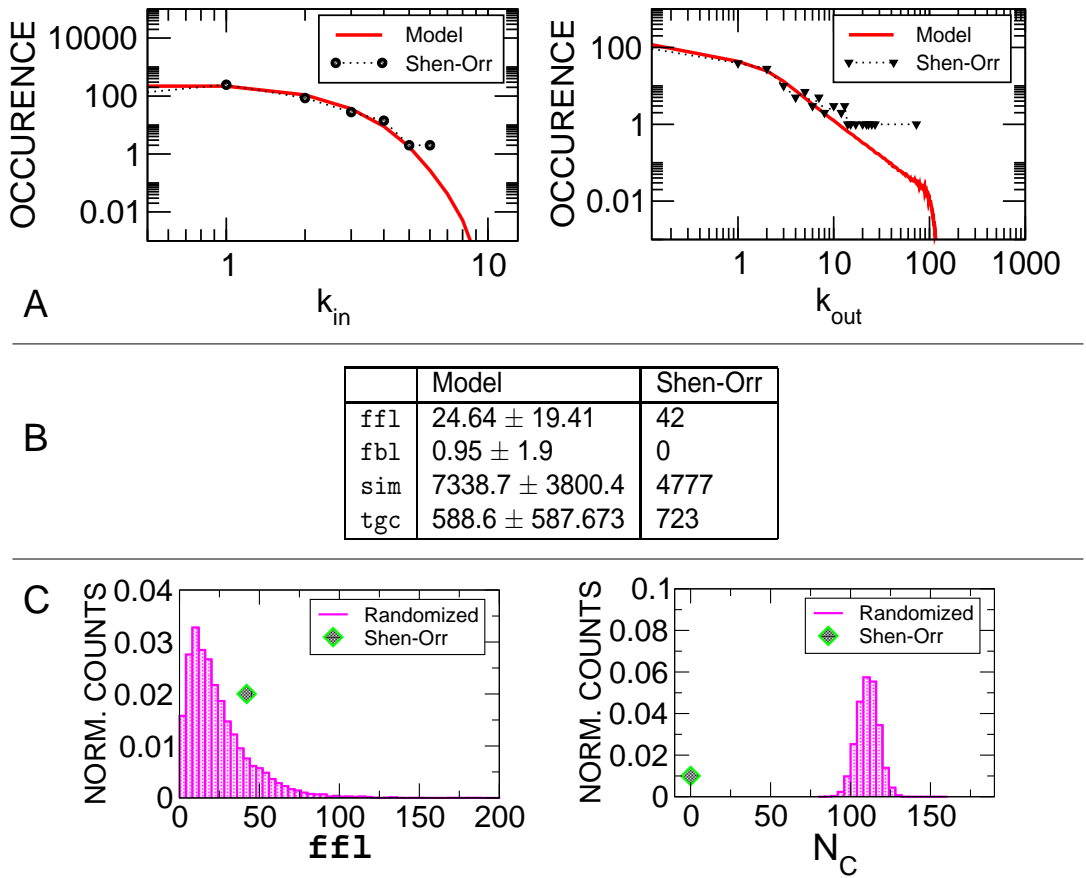}
  \caption{Application of the model to the Shen-Orr dataset. Example
    of fit and observed features. The plots refer to the parameter set
    $\beta = 1.83, \alpha = 0.5, m_n/n = 0.2766$, with a cutoff on the
    maximum outdegree at $18\%$ of the nodes as described in the text.
    (A) in- and out-degree histograms of the empirical graph, compared
    to the random ensemble. While the tail of the outdegree may not
    seem a good fit, we note that the integral of the $> 13$ tail, or
    the estimated number of ``global regulators'', of the two laws are
    remarkably similar (8 in the empirical graph vs 9.7 in the
    randomized network) so that this has to be regarded as a good
    agreement. (B) Table comparing the subgraph content (for the
    three-node subgraphs analyzed here) of the model graphs with the
    empirical one. The two quantities are in general very similar,
    with the exception of the \texttt{ffl}, which deviates from
    average, but only slightly. (C) The feedback in the graph deviates
    from average more than the triangular subgraphs. Left panel: the
    distribution of \texttt{ffl}s compared with the empirical value.
    Right panel: the feedback of the random and empirical graphs
    differ.  $N_C$ Measures the number of nodes left in a graph after
    pruning the input and output trees, as described
    in~\cite{gammachi}.}
  \label{fig:fit}
\end{figure*}

The ensemble obtained with this procedure fits quite well the
empirical in- and out-degree sequences (Fig.~\ref{fig:fit}A), Also, the
model reproduces the empirical number of transcription factors, roots
and leaves as average values. 
As a remark, we note that, unless new prescriptions for the generation
of the graphs, and thus new parameters, are introduced, roots, leaves
and transcripiton factors cannot be reproduced well with smaller
values of $\beta$ than the ones we used. One can take this as a
confirmation that the range of values for the exponent obtained with
our fitting procedure are reasonable.

We also measured the three-node subgraph content of the null model and
compared it with the empirical data and the model ensemble are very
close (Fig.~\ref{fig:fit}B). The only exception is the \texttt{ffl},
with a slight deviation, that, however, is much less significant than
with the degree-conserving ensemble. Thus, in term of these
observables, one obtains similar graphs as the empirical one. This
means that in the resulting ensemble the average motif content can be
regarded as an invariant, rather than as an observable.
Finally, we quantified the feedback properties (Fig.~\ref{fig:fit}C).
In order to do this, we measured the number $N_C$ of nodes left in the
graph after pruning its input and output tree-like components with an
iterative decimation algorithm~\cite{gammachi,pnas}.
In particular, none of the graphs we generated was treelike and
feedforward as the empirical one. One may then speculate that the
motif content and the hierarchical properties, two important
properties are somehow related.

\section{Conclusions.}

We presented an algorithmic way to generate directed graphs with,
asymptotically, power-law outdegree and compact indegree, easily generalizable
to different kinds of graphs. The discussion was carried out having in mind an
application in the realm of transcription networks, although there are many
possible connections with other experimentally accessible complex networks,
including biological ones.
Compared to other techniques, our model has the advantage to be quick in
generating large graphs, as it is not designed to preserve a prescribed degree
sequence, but rather to generate an ensemble with given degree distributions.
As such, it is an interesting tool to characterize topological observables in
large graphs. Most importantly, many of the relevant observables are
accessible analytically, for any value of $n$. We supplied here as an example
the evaluation of the mean number of subgraphs, roots and leaves and hub.

We should add here a comment regarding the is not evident that the proposed
approach is more efficient that the Molloy-Reed algorithm~\cite{MR95}, which
generates ``stubs'' with desired in- and outdegree sequences, and matches the
stubs to generate the graphs.  This model could be re-cast to be similar in
spirit, in the sense that one could fix the relevant distributions depending
on parameters, ad throw the degree sequences from the distributions. Once the
number of connections for each node has been drawn from the expected degree
distribution and without avoiding multiple connections, the computational cost
of pairing the stubs is order E (number of edges), so in sparse networks this
could be less than order $n^2$, and in non-sparse networks it could be $n^2$.
Despite the algorithm suffers from the undesired production of multiple edges,
due to the computational complexity of pairing hubs, for a compact indegree
distribution, this computational cost can be small~\cite{CFS05}, allowing a
practical applicability in some regimes. On the other hand, we think that our
approach remains competitive, as its computational cost is not affected by the
complexity of the graph ensemble, and, as we have shown, is very versatile for
analytical calculations.

Regarding the subgraph structure, we note that while \texttt{ffl}s always
dominate on \texttt{fbl}s, there are qualitatively different behaviors
depending on the exponent $\beta$.  The most marked dominance is found for
smaller values of $\beta$, and is further increased by considering rectangular
matrices (i.e. asymptotically compact indegree). Thus, the degree distribution
poses some important constraints on the dominant subgraphs in our null model.
We would like to spend a few more words on these scaling laws with system size
$n$. In our model the scaling of $\mu_n$ with the decay exponent $\beta$
pilots the transitions of all the observables. In particular, it renders
necessary to consider rectangular matrices to obtain an asymptotically compact
indegree if $\beta \le 2$.
This behavior is interesting on theoretical grounds, and shows how much the
distributions for the in- and outdegree in transcription networks are strongly
unbalanced. For example, in the model described in section~\ref{sec:var},
where the indegree is allowed to have a power-law tail, the situation is
rather different.
In the case of transcription networks, there is an observed scaling law of the
fraction of transcription factors (nodes that have at least one outgoing
link).  This is a power-law $n^{\zeta}$~\cite{vN03} with positive exponent
$1<\zeta<2$. Looking at the distribution of roots, one easily realizes that
this behavior forbids any asymptotic limit assuming the graph structure of our
model, and is thus incompatible with it. At the light of our calculations, we
can observe that it is likely that for larger values of $n$ the outdegree
ceases to follow a power-law, and/or the average indegree ceases to be finite,
the opposite trend to that observed in small networks.
Experimental observations of larger transcription networks will elucidate this
question.

We should stress here that the above considerations apply mainly to the model.
Nevertheless, we showed that in principle our model can be used for
direct statistical comparisons, as an alternative to the more
stringent constraint of preserved degree sequences.
An example of such a fitting procedure, produced an ensemble of
networks that resemble the empirical one of E.~coli in terms of degree
distribution, number of links, roots, leaves and transcrption factors.
Interestingly, the null ensemble produced this way also has a very
similar three-node subgraph content as the empirical graph. On the
other hand, the feedback properties are very different.
The outcome of such a comparison might depend on the invariance
criteria used for the fitting. This is an interesting feature that can
be used to produce flexible null models, depending on the quantities
of interest. On the other hand, this feature makes the handling of the
model more delicate than the standard degree-conserving
randomizations. In particular, a more exhaustive analysis than that
presented here is needed to draw clearcut conclusions on experimental
graphs~\cite{in_prep}. 
Clearly, the degree sequences of, for example, the E.~coli network,
are not stringently fixed by any physical of biological constraint.
Rather, the network, during evolution (and within a population), moves
in a larger ``space of possible interactions'', determined by
selective pressure and other biological constraints, which has not
been strictly identified yet.  Generalizations of our null model might
help exploring this evolutionary problem.

Finally, we showed how the coin-toss algorithm, or exchangeable graph model,
has a wider range of application than the main example examined here. To
illustrate this, we explained how, with the same technique, one can obtain
directed and undirected power-law random graphs.  Obviously, the range of
possibilities is even larger if one starts to play with the probability
distribution for the biases $\pi_{n}(d\theta)$.
For this reason, on more abstract grounds, the model can be useful in the
context of the theory of correlated random networks~\cite{MS05,BL02}. It is a
quick algorithm easy to implement and to analyze theoretically. Indeed,
because of its simple formulation, the potential for further analytical
calculations is large. For example, one can evaluate the kernel of $A$, which
is useful in connection with problems of the Satisfiability class, which have
seldom been analyzed on non-Poisson random
graphs~\cite{Kol98,Lev05,MPZ02,LJB05}.

\appendix

\section{Average of \texttt{ffl}}
\label{sec:fflav}

This appendix reports in more detail the calculation of the mean number of
\texttt{ffl}s for $1<\b<2$. Starting from the definition, we obtain with
straightforward calculations
\[
\begin{split}
&<\CN_\mathtt{ffl}(G_n)> \\ &= 6 {n \choose 3}
\frac{(\b-1)^2}{[(n/\a)^{\b-1}-1]^2 }  
\int_{\a/n}^1 \th^{2-\b} d\th \int_{\a/n}^1 \th^{1-\b} d\th \\
&=
\frac{\a^{2\b-2}(\b-1)}{(3-\b)(2-\b)} n^{5-2\b} \left[1-\frac{3}{n}
+\frac{2}{n^2}\right]\\
&\times \left [1-\left 
(\frac{\a}{n}\right)^{3-\b}-\left(\frac{\a}{n}\right)^{2-\b}+
\left(\frac{\a}{n}\right)^{5-2\b} \right] \\
\end{split}
\]
Note that, since the finite $n$ formula for the mean is known exactly, the
finite size scaling can be computed analytically, simply by isolating the
leading terms in the approach to the asymptotic limit. 
For example, in the
case of the \texttt{ffl} average computed above, one has
\[
\begin{split}
<\CN_\mathtt{ffl}(G_n)>&=
\frac{\a^{2\b-2}(\b-1)}{(3-\b)(2-\b)} n^{5-2\b} \\
\times &\left [1-\left 
(\frac{\a}{n}\right)^{2-\b}+o\left 
(\frac{1}{n^{2-\b}}\right)\right ] .\\
\end{split}
\]


\section{Variance of \texttt{fbl} vs \texttt{ffl}.}
\label{sec:A1}
 
We report here the calculation of the standard deviation of
feedforward and feedback loops, in the case of square matrices.
The key point is to evaluate $\langle\CN_{\mathtt{ffl}}(G_n)^2
\rangle$ and $\langle \CN_{\mathtt{fbl}}(G_n)^2\rangle$. Again, for
the sake of simplicity, we will deal only with square matrices. It is
clear that $\langle\CN_{\mathtt{fbl}}(G_n)^2 \rangle= \sum_{t \in
  \tau} \sum_{s \in \tau} P\{ s,t \in G_n \} $, $\tau$ being the set
of all feedback loops contained in the complete $n$ graph.
Analogously one obtains $\langle \CN_{\mathtt{fll}}(G_n)^2\rangle$
taking as $\tau$ the set of all feedforward loops. Simple 
calculations give
\[  
\begin{split}
\langle \CN_{\mathtt{fbl}}(G_n)^2\rangle&=
4{n \choose 3}{n-3 \choose 3}\mu_n^6 +
12  {n \choose 3}{n-3 \choose
  2}\mu_n^2
\delta_{2,n} \\
&+6(n-3) {n \choose 3}(\mu_n^3+\mu_n^2\delta_{2,n}^2) +2{n \choose
  3}\mu_n3 \\
\end{split}
\]
where $\delta_{i,n}:= \int_0^1 \th^i\pi_n(d\th)$. Hence, one obtains
\[
Var(\CN_{\mathtt{fbl}}(G_n))
 \sim 
\left\{%
\begin{array}{ll}
n^{5(2-\b)}  &\,  \, \,  \,\mathrm{if} \ \  1 < \b < 2  \\
(\log n)^4 &\,  \, \,  \,\mathrm{if} \ \ \b = 2 \\
\frac{1}{3}(\a\frac{\beta-1}{\beta-2})^3 &  \,  \, \,  \mathrm{if} \ \
\b \geq 2 \\
\end{array}%
\right.
\]
As for $\CN_{\mathtt{ffl}}$, the computations are longer, but
essentially the same. The problem is that $P\{ s,t \in G_n \}$ can
take many different expression depending on $s$ and $t$. 
With some simple but tedious calculations one gets
\[
\begin{split}
<\CN_{\mathtt{ffl}}(G_n)^2>&= 6 {n \choose 3} A_n
+ 6(n-3) {n \choose 3} B_n \\
&+ 12 {n \choose 3}{n-3 \choose 2}C_n +36{n \choose 3}{n-3 \choose 3}D_n\\
\end{split}
\]
with
$A_n=\delta_{1,n} \delta_{2,n}
+ \delta_{2,n}^2 +\delta_{1,n}^2 \delta_{2,n}$,
$B_n=\delta_{2,n} \delta_{3,n}
+ 5 \delta_{1,n}\delta_{2,n}^2 +3 \delta_{1,n}^2 \delta_{3,n}
+ \delta_{3,n}^2+2\delta_{1,n} \delta_{2,n} \delta_{3,n}
+2 \delta_{2,n}^3+4 \delta_{1,n}^2 \delta_{2,n}^2 $,
$C_n=2\delta_{1,n} \delta_{2,n}\delta_{3,n}+
+ \delta_{1,n}^2 \delta_{4,n} + 5 \delta_{1,n}^2 \delta_{2,n} +
\delta_{2,n}^3 $ and 
$D_n=\delta_{1,n}^2 \delta_{2,n}^2$.
Hence, 
\[
\begin{split}
Var(\CN_{\mathtt{ffl}}(G_n))&= 6 {n \choose 3} A_n
+ 6(n-3) {n \choose 3} B_n \\
&+ 12 {n \choose 3}{n-3 \choose 2}C_n -36 R_n D_n\\
\end{split}
\]
with $R_n=[{n \choose 3}-{n -3 \choose 3}]$.
For example, if $2 <\beta<3$, the last expression gives 
\[
Var(\CN_{\mathtt{fll}}(G_n)) \sim n^{2(\beta+1)}.
\]
%

%


\end{document}